\documentclass[aps,prd,final,twocolumn,floats,floatfix,nofootinbib,10pt]{article}

\usepackage{blindtext}  
\usepackage{graphicx}
\usepackage{amsmath}
\usepackage{amsfonts}
\usepackage{amssymb}
\usepackage{physics}
\usepackage{amstext}
\usepackage[english]{babel}
\usepackage{helvet}
\usepackage{microtype}
\usepackage{hyperref}
\usepackage{url}
\usepackage[margin=0.75in]{geometry}

\begin{document}

\twocolumn[
\begin{@twocolumnfalse}
\title{Two sites coherence and visibility}

\author{Abanoub Mikhail}

\date{\today}
\maketitle \thispagestyle{empty}

\begin{abstract}
Wave-particle duality and the superposition of quantum mechanical states furnish quantum mechanics with unique features which distinguishes it from classical mechanics and give it the apparently counter-intuition interpretation. The two principles are responsible for the observation of the interference effects of quantum particles such as electrons, atoms and molecules. Visibility is a measure of the wave nature and can be though of as a "normalized" coherence quantifier. Reduction in the visibility arises from dephasing (decoherence): a process in which the relative phases get partially or totally destroyed leading to domination of the particle nature. We calculate the coherence and visibility of an ensemble of a single electron and two electrons on two sites using the density matrix formulation. For such a system of electrons, visibility and predictability (a particle nature measure) follow an oscillatory-complementary temporal evolutions and neither of them depends on the single particle energy.
\end{abstract}

\end{@twocolumnfalse}]

\section{Introduction}
The superposition principle is a fundamental concept in physics. It applies to all kinds of waves: mechanical and electromagnetic in classical mechanics as well as De Broglie waves in quantum mechanics. The superposition principle states that if a quantum system can be found in a state $\psi_1(t)$ \footnote{Variables such as position and spin are included implicitly} and also in a state $\psi_2(t)$, then it can be found in any linear combination of them \footnote{we have absorbed all constants into $\psi_1(t)$ and $\psi_2(t)$}  $\psi(t)=\psi_1(t)+\psi_2(t)$\footnote{ The extension to more than two states is straightforward}.

In the position representation, when one evaluates the probability density $|\psi(t)|^2$, in adition to the separate probabilities of the two waves $|\psi_1(t)|^2$ and $|\psi_2(t)|^2$, there are cross terms $\psi^*_1(t) \psi_2(t)+\psi^*_2(t) \psi_1(t)$ which is called the interference terms as they are responsible for the observation of the interference effects \footnote{In the interference experiments such as double slits experiment, the interference pattern is defined as the response of the fixed detectors as a function of phases \cite{Marquardt2008}}. 

When light from the sun passes through a hole, we do not see diffraction pattern and when it passes through two slits, there is also no interference pattern. That is because sun light is \textbf{incoherent} \footnote{its constituent photons do not maintain constant relative phase and it has different frequencies as well.}. Therefore, incoherent sunlight behaves as particles. Once it has been made coherent, diffraction and interference patterns reappear. Thus, coherence is a manifestation of the wave nature. The same is true for quantum particles. In a \textbf{coherent superposition}, the interference effects can be observed due to the existence of interference terms while the interference terms vanish --or average out to zero-- in \textbf{incoherent superposition} \cite{Gennaro2001}. For example, consider the state
\begin{equation}
\psi(t) = \psi_{1}(t)+\psi_{2}(t)
\label{2.1}
\end{equation} 
the probability distribution is given by 
\begin{eqnarray}
{|\psi|}^2&=&{|\psi_1|}^2+{|\psi_2|}^2+\psi^*_{1} \psi+\psi^*_2 \psi_1  \nonumber   \\
            &=&{|\psi_1|}^2+{|\psi_2|}^2+2{|\psi_1 \psi_2|}\cos\left(\alpha_{12}\right)
\label{2.3}
\end{eqnarray}
where the last equality follows from writing $\psi_1$ and $\psi_2$ in the phasor form as $\psi_1=|\psi_1|\exp\left(i\alpha_1\right)$, $\psi_2=|\psi_2|\exp\left(i\alpha_2\right)$ and the relative phase is defined by $\alpha_{12}=\alpha_1-\alpha_2$. A relative phase difference of $\frac{\pi}{2}$ completely destroys the interference term. That is referred to as destructive interference \footnote{A relative phase difference of $\frac{\pi}{2}$  implies orthogonality between $\psi_1(t)$ and $\psi_2(t)$}. The superposition is called incoherent in this case. A maximally coherent superposition occurs when $\alpha_{12}=0$. The relation between the coherence of the superposition state and the coherence of the two terms being imposed is given in the form of upper and lower bounds as discussed in \cite{Qiu2016} 

Conventionally, visibility is a measure of the contrast of the interference pattern. For interference experiments with light such as the double slit experiment , the fringe visibility is defined locally as \cite{Wolf1999}
\begin{equation}
\textrm{V}=\frac{I_{max}-I_{min}}{I_{max}+I_{min}}
\label{Eq1.1}
\end{equation}
$I_{max}$ is the maximum intensity of light and $I_{min}$ is the minimum light intensity. Classically, Light intensity is proportional to the square of the electric field amplitude. This definition of visibility is used as a measure of the contrast of the interference pattern.

Much work has been conducted on finding a wave and particle nature measures \cite{Bimonte2003, Bimonte2003_2}.  Various wave particle complementarity relations has been proposed. Jaeger et al found a duality relation between the distinguishably of the path of propagation of a particle and the fringe visibility when the amplitudes of two paths are combined. A second complementary relation relates the visibility of one-particle interference fringe to the visibility of two-particle interference fringes were also found by the same authors \cite{Jaeger1995}. If an experimentalist have complete knowledge about the path that the quanton take, the interference pattern gets destroyed. That implies that the quantum particles behaved as classical particles in accordance with Bohr's principle of complementarity. For a two beam interferometer, a useful measure of the particle properties is the predictability $P$. The predictability reflects our ability to predict the path(beam) the quanton takes(in). It is defined as $P=|\rho_{11}-\rho_{22}|$ where $\rho_{ii}$ for $i=1,2$ are the diagonal elements of the density matrix. A wave-particle duality relation was introduced as \cite{Yasin1988}
\begin{equation}
P^2+V^2 \le 1
\end{equation}
The equality holds for maximally coherent states and the inequality for partially coherent states. The predictability was extended to $n$-path interferometer in Ref. \cite{Durr2001}.  Readers who find it helpful to study coherence and visibility in the framework of double and multipath interferometers, may find this preprint useful \cite{Tabish2019}.

The main objective of this paper is to investigate the temporal evolution of coherence and site predictability in a simple electronic system other than the interferometers' experiments though these experiments are used to illustrate most of the concepts. Hubbard model of two electrons on two sites provides a perfect example as it incorporate the essential physics of delocalized-electrons in solids --the limit of weak electron-electron interaction--  and localized-electrons in molecules --the limit of strong electron-electron interaction. While delocalization is a wave property, localization is a particle property. Hence, the significance of visibility and site predictability for this system. In addition, a scaling of coherence is suggested which, creditably, is independent of the dimensionality of the Hilbert space. 

This paper is organized as following: a quantifier of coherence is introduced in subsection 2.2. In the last part of section 2, the reader is introduced to the process of dephasing in which the relative phase shift is randomized followed by partial or total destruction of the interference effects, hence, a reduction of the visibility and the transition of the system from the quantum like to the classical like behavior. In section 3, visibility, as a measure of the wave nature, is defined and linked to a coherence quantifier. Starting from section 4, we apply the concepts and techniques developed in previous sections to Hubbard model of single and two electrons on two sites.

\section{Theoretical basis}
\subsection{The density matrix}
 Although the state vector contains the maximal information about the system, in many situations, the system is not isolated and full information about it is not known. This loss of information is the reason behind the introduction of the density matrix in place of the state vector.  Pure states refer to states in an ensemble that is prepared in the same Eigenstate or in the same superposition of Eigenstates. Mixed states refers to states in an ensemble in different superposition of Eigenstates --in a statistical mixture of pure states.Therefore, the quantum mechanical ensemble is characterized by the statistical operator (also called the density matrix operator) $\rho$ defined by
\begin{equation}
\rho=\sum_{m}P_m\ket{\psi_m}\bra{\psi_m}
\label{3.1}
\end{equation}
where $P_m$ is the probability to find the system in the pure state $\ket{\psi_i}$. In terms of of a complete orthonormal basis $\ket{i}$, the pure state is $\ket{\psi_m}=\sum_{i}c_i^m\ket{i}$ and the density matrix becomes 
\begin{equation}
\rho=\sum_{i,j}\left(\sum_{m}P_m c_i^{(m)}( c_j^{(m)})^*\right)\ket{i}\bra{j}
\label{3.2}
\end{equation}
It follows that the density matrix elements are given by 
\begin{equation}
 \rho_{ij}=\sum_{m}P_m c_i^{(m)}( c_j^{(m)})^*
 \label{3.3}
\end{equation}
The diagonal elements give the probability of finding the system in one of the basis states $\ket{i}$, therefore, adding up these probabilities must be one. Meanwhile, the off-diagonal elements are a measure of the coherence between different states of the system --see the next section.$\textrm{Tr}(\rho^2)$ can be used to test whether a given density matrix describes a pure or a mixed state as it is equal to one for pure states and less than one for mixed states. For more details on the density matrix and its applications, the reader is directed to the book by Karl Blum \cite{Blum2012}.

\subsection{Quantifying quantum coherence}
A rigorous and reliable quantifier of coherence was proposed in Ref. \cite{Baumgratz} by Baumgratz et al. In this approach, for a fixed basis $\ket{i}_{i=1,...,n}$ of n-dimensional Hilbert space \texttt{H}, all density matrices which are diagonal in this basis are called incoherent. This set of quantum states is labeled by $I \subset \texttt{H}$. Therefore, all density operators $\rho^{I} \in I$  take the form $\rho^I=\sum_{i=1}^{n}\rho^I_i\ket{i}\bra{i}$ . A maximally coherent state is given by $\ket{\psi_{n}}=\frac{1}{\sqrt{n}}\sum_{i=1}^{n}\ket{i}$ and the coherence of this state is used as a unit of coherence.

Any coherence measure $C$ should satisfies the following properties: \cite{Baumgratz} (1) it must vanishes on the set of incoherent states $C(\rho^I)=0$ (2) Monotonicity under under incoherent completely positive and trace preserving maps (3) Nonincreasing under mixing of quantum states (convexity). The $l_1$ norm of the density matrix satisfies the previous conditions and is chosen as a suitable measure of coherence. Hence the coherence becomes 
\begin{equation}
C(\rho)=\sum_{i\not= j}{|\rho_{ij}|}
\label{3.4}
\end{equation} 
 
\subsection{Dephasing (Decoherence)}
As pointed out in,\cite{Buks1998} the complementary principle prevents a perfect knowledge of conjugate pairs of physical quantities simultaneously. Wave particle duality is one of such pairs. The partial or total destruction of the interference terms yield a reduction in the visibility. This process of loss of the coherence is called decoherence or dephasing -- the existence of coherence lies in the observation of interference effects.\cite{Biswas2017}  In the double slit interference experiment, any attempt of the experimentalist to measure any property of the interfering quantons leads to weakening followed by partial or total destruction of the interference pattern. That is the interference pattern is visible --hence the wave-like nature dominates-- when we do not know exactly the path that the quantons take and when there is no leakage of information about the them to the environment, an observer or any measuring instrument.

The study of dephasing can be proceed by realizing the changes that the system leaves on the environment or by considering the randomization of the system relative phase \cite{Ady1990} due to the existence of the environment. In the latter description, the path that a particle takes becomes uncertain. A paper by Yakir Aharonov and his colleagues illustrates this point and they proved the equivalence of both approaches.\cite{Ady1990} Their argument starts as follow: assume a left wave packet $\ket{r(x,t)}$ and a right wave packet $\ket{r(x,t)}$ crossing a ring in two opposite directions. Their interference is examined after they travel one-half of the ring in two opposite direction. The right wave packet can interact with the environment while the left wave packet cannot. The interaction Hamiltonian is assumed to have the following form
\begin{equation}
H_{int}=V(x,\eta)
\label{3.5}
\end{equation}
where $x$ is the particle space coordinate and $\eta$ is the environment coordinates. The initial wave function is $\left[\ket{l(x)}+\ket{r(x)}\right]\otimes\ket{\chi(\eta)}$. After time $t=\tau$, the wave function becomes
\begin{align}
\ket{\psi(\tau)}&=\ket{l(\tau)}\otimes\ket{\chi(\eta)} \nonumber \\
                         &+\ket{r(\tau)}\otimes\exp\left(\frac{-i}{\hbar}\int^{\tau}_0V(x_r(t),\eta)\,dt\right)\ket{\chi(\eta)}
\end{align}
The interference term is given by \\
$2\textrm{Re}[l^*(x,\tau)r(x,\tau)$ \\
$\times\bra{\chi(\eta)}\exp\left(\frac{-i}{\hbar}\int^{\tau}_0V(x_r(t),\eta)\,dt\right)\ket{\chi(\eta)}]$\\
\\
Thus the effect of the environment on the particle is encoded in the factor $D(\phi,\eta)=\bra{\chi(\eta)}e^{i\phi_(\eta)}\ket{\chi(\eta)}$  where $\phi(\eta)\equiv \exp\left(\frac{-i}{\hbar}\int^{\tau}_0V(x_r(t),\eta)\,dt\right)$ is the phase shift. The first interpretation of the authors is based on the previous equation: quantum interference is lost when the two interfering waves shift the environment into two orthogonal states.

 In the second interpretation, 
\begin{eqnarray}
D(\phi,\eta)&=&\int {|\chi(\eta)|}^2 \exp\left(\phi(\eta)\right) d{\eta} \nonumber \\\
                   &=&\int d{\phi} {|\chi(\eta)|}^2 \exp\left(\phi(\eta)\right)\frac{d\eta}{d\phi} \nonumber \\
                   &=&\int P(\phi)\exp\left(\phi(\eta)\right) d\phi 
\label{D2.4}     
\end{eqnarray}
 where $P(\phi) \equiv \abs{\chi(\eta)}^2 \frac{d\eta}{d\phi}$ is the distribution function. Since Equation (\ref{D2.4}) is nothing but the stochastic average of the relative phase shift i.e, $\langle e^{i\phi}\rangle_{\phi}=\int P(\phi)\exp\left(\phi(\eta)\right) d\phi $. In this approach, $\phi$ is not well-defined. Rather it becomes a statistical variable described by the distribution function $P(\phi)$. \cite{Ady1990} If the factor $\langle e^{i\phi}\rangle_{\phi}$  vanishes when averaged over all the environment degrees of freedom, the interference terms become zero and the system behaves classically. 

There are also other dephasing-inducing processes such as scattering and excitation. Dephasing can be thought of a result of: a. Stochastic classical process b. Interaction of the system with a bath in a random initial state c. Quantum randomness, see Ref. \cite{Marquardt2008} for further detail.

\section{Defining and linking visibility to coherence}
Apparently the factor $\langle e^{i\phi}\rangle_{\phi}$ describes the modifications to the interference term due to the existence of the environment. Hence, the absolute value of this factor can be used more or less as a measure of the visibility.\cite{Marquardt2008} If one insists on using the extrema of the interference terms, then, we should replace the concept of light intensity by the squared modulus of the  probability amplitude and rewrite the visibility as 
\begin{equation}
V=\frac{|\psi^C|^2-{|\psi^I|}^2}{|\psi^C|^2+|\psi^I|^2}
\label{2.4}
\end{equation}
If  ${\psi}^C$ is  maximally coherent states, constructive interference is possible --the two states are in phase. Meanwhile, an incoherent state $\psi^I$ exhibits no interference $|\psi^I|^2={|\psi_1|}^2+{|\psi_2|}^2$ because the two states are $\frac{\pi}{2}$ out of phase --pure particle behavior. 
Even if the basis states are orthonormal, one may think of other situations where interference is natural to the problem. For example, 
Suppose a two state system in the initial state 
\begin{equation}
\ket{\psi_{i}(t)}=a_{1}(t) \ket{1}+a_{2}(t) \ket{2}
\label{3.13}
\end{equation}
and the final state of the system is 
\begin{equation}
\ket{\psi_{f}(t)}=b_{1}(t) \ket{1}+b_{2}(2) \ket{2}
\label{3.14}
\end{equation}
The transition amplitude from the initial to the final is given by 
\begin{align}
{|\bra{\psi_{f}(t)}\ket{\psi_{i}(t)}|}^2&={|a_1|}^2{|b_1|}^2+{|a_2|}^2{|b_2|}^2 \nonumber \\
                                                              &+2\Re(a_{1}^{\ast}a_2 b_1 b_2^{\ast})
\end{align}
In the phasor form $a_i(t)={|a_{i}(t)|} e^{\alpha_i(t)}$ and $b_i(t)={|b_{i}(t)|} e^{\beta_i(t)}$ for $i=1,2$ where the relative phase is defined by $\alpha_{12}=\alpha_1(t)-\alpha_2(t)$ and $\beta_{12}=\beta_1(t)-\beta_2(t)$, the interference term is
\begin{equation}
2\Re(a_{1}^{\ast}a_2 b_1 b_2^{\ast})=2{|a_1b_1a_2b_2|}\cos\left(\alpha_{12}-\beta_{12}\right)
\label{3.15}
\end{equation}
it is reduced by a factor of $\cos\left(\alpha_{12}-\beta_{12}\right)$. Again, the interference term completely vanishes when $\alpha_{12}-\beta_{12}=\frac{\pi}{2}$. It follows from equation (\ref{2.4}) that the visibility for this system is 
\begin{equation}
V(t)=\frac{{|a_1b_1a_2b_2|}}{{|a_1|}^2 {|b_1|}^2+{|a_2|}^2 {|b_2|}^2+{|a_1b_1a_2b_2|}}
\label{3.16}
\end{equation}

 Finally we extent the definition of visibility to the case of mixed states. In a paper by Stephan D\''urr, the author proposed criteria for the visibility to be a good measure of the wave properties:\cite{Durr2001}\\
(1) It should be possible to give a definition of V that is based only on the interference pattern without explicitly referring to the matrix elements of $\rho$\\
(2) V should vary continuously as a function of the matrix
elements of $\rho$.\\
(3) If the system shows no interference V should reach its global minimum\\
(4) If $\rho$ represent a pure state and all the states are equally populated, V should reach its global maximum. \\
(5) V considered as a function in the parameter space $(\rho_{11},\rho_{12},...,\rho_{nn})$ should have only global extrema, no local ones.  \\
(6) V should be independent of our choice of the coordinate system.\\
Notice that conditions (3) and (4) are analogous to the conditions imposed on the coherence quantifier. A straight forward extension of Equation (\ref{2.4}) to mixed states yields 
\begin{eqnarray}
V(\rho)&=&\frac{|\rho^C|-|\rho^I|}{|\rho^C|+\rho^I|}   \label{3.17}\\
           &=&\frac{0.5\sum_{i\not=j}|\rho_{ij}(t)|}{\sum_{i} \rho_{ii}(t)+0.5\sum_{i\not=j}|\rho_{ij}(t)|} 
\label{3.17}
\end{eqnarray}
 where $\rho^C$ is a coherent state. When the density matrix $\rho^C=\rho^I$, $V(\rho)$ is zero and it approaches unity as the sum of the off-diagonal elements gets very large. All the conditions satisfied by the coherence quantifier are also satisfied by $V(\rho)$. In this sense, Equation (\ref{3.17}) is nothing but a normalized coherence measure.

\section{Applications}
\subsection{The visibility of two and four path interferometers}
Now the visibility is defined as a measure of the wave properties where its zero was assigned to a point of pure particle behavior. The visibility was scaled in Ref. \cite{Qureshi2015} as 
\begin{equation}
V_C(\rho)=\frac{1}{n-1}\sum_{i \not=j}\rho_{ij}
\label{4.1}
\end{equation}
where $n$ is the dimensionality of the Hilbert space. The factor $(n-1)$ is not arbitrary. It is actually equal to the coherence of a maximally coherent states of $n$ dimensions. In the extreme limits, $ V_C$ is $0$ for incoherent states and $1$ for maximally coherent states. Meanwhile, $V$ and $V_C$ do agree for incoherent states and $V$ approaches $V_C$ in the limit of very large $C(\rho)$, they do not coincides in between. $V$ can be used to compare the wave properties domination in different experiments. For example, we are expecting that the wave properties will become more and more dominant as the number of slits increase i.e, the interference fringes become more bright and sharp. That is because as the number of slits increases, the interference terms become very large and so does the coherence; hence, $V$ approaches unity.

To illustrates this point, let us assume two path interferometer and four path interferometer where the quantons are prepared in maximally coherent states in both experiments. If there is no path-detection or decoherence involved, the density matrices can be written respectively as 
\begin{equation}
\rho=\frac{1}{2}\begin{bmatrix}
1&1\\
1&1\\
\end{bmatrix}
\end{equation}
\begin{equation}
\rho=\frac{1}{4}\begin{bmatrix}
1&1&1&1\\
1&1&1&1\\
1&1&1&1\\
1&1&1&1\\
\end{bmatrix}
\label{4.2}
\end{equation}
while $V_C$ is $1$ for both experiments, $V$ is $\frac{1}{3}$ and $\frac{3}{4}$, respectively i.e, increases with coherence. In general, for $n$-slit experiment prepared in maximally coherent states, $V$ is equal to $\frac{\frac{n}{2}(n-1)}{2+\frac{n}{2}(n-1)}$ which indeed tends to $1$ as $n\rightarrow \infty$
\subsection{Hubbard Model}
We have used the interference experiments to illustrate most of the concepts, but the formulas we have derived does not assume a particular system.
\subsubsection{The Hamiltonian}
 Despite its simplicity, Hubbard Model \cite{hubbard1963} captures the essential physics of electrons in atoms, molecules and solids \cite{schmidt1987}. The model has been applied to the theory of magnetism , Mott metal-insulator transition \cite{mott1968,mott1990,raghu2008} and high temperature superconductors. In solids, electrons interact with each other via screened Coulomb potential. When the site (the atom) is vacant or has a single electron, the electron-electron interaction energy is zero. When it is paired with opposite spin electrons, the interaction energy is assigned the parameter U \cite{Lancaster2014}. That is electrons are correlated with each other in the same site but weakly correlated in different sites \cite{hubbard1963}. By hopping between site $i$ and site $j$, electrons save kinetic energy $T_{ij}$ \cite{Lancaster2014}. Taking the interaction of electrons with nuclei into account, the Hamiltonian under the previous assumptions is 

\begin{equation}
\hat{H}=-\sum\limits_{i\neq{j}\sigma}T_{ij}c^{\dagger}_{i\sigma}c_{j\sigma}+\sum\limits_{i{\sigma}}\epsilon_{i}n_{i{\sigma}}+U\sum\limits_{i}n_{i\uparrow}n_{i\downarrow}
\label{4.3}
\end{equation}

where $c^{\dagger}_{i\sigma}$ creates an electron at position $i$ with spin $\sigma$ (either $\uparrow$ or $\downarrow$) and $c_{j\sigma}$ destroys a $\sigma$-spinned electron at position $j$. The sums over $i$ and $j$ are understood to be over different sites ($i\neq{j}$) and no spin flopping processes are allowed. $\epsilon_i$ is the one-electron energy in site $i$. The operators $n_{i\uparrow}$ and $n_{i\downarrow}$ count the number of up-electrons and the number of down-electrons, respectively. Note that the hats have been dropped from all operators for clarity. The model in this form  successfully accounts for localization of electrons in molecules and their delocalization in metals.\cite{alvarez2001} The extent of delocalization, wave property, is quantified by the visibility.
\subsubsection{One electron on two sites}
Assume an electron on two sites. Let $\ket{1}$ and $\ket{2}$ represent spatially-localized electron states at sites $1$ and $2$, respectively. If the electron adapts pure particle behavior, its state is either $\ket{1}$ or $\ket{2}$ at a particular time instant. However, quantum mechanics allows the electron to be in any linear superposition of $\ket{1}$ and $\ket{2}$ simultaneously. Since the two sites are indistinguishable, the wave-like behavior, hence interference, is possible \cite{Buks1998}.

The general state of an electron on two sites can be written in the following form 
\begin{eqnarray}
\ket{\psi(t)}&=& \phi_{1}(t)\ket{\uparrow;0}+\phi_{2}(t)\ket{0;\uparrow} \nonumber \\
             &=& \phi_{1}(t)\ket{1}+\phi_{2}(t)\ket{2}           
\end{eqnarray}
where the states: $\ket{1}=\ket{\uparrow;0}$ and $\ket{2}=\ket{0;\uparrow}$. 


The general solution is a linear superposition of the Eigenvectors $v_1$ and $v_2$ that have Eigenvalues $E_1=\epsilon-T$ and $E_2=\epsilon+T$, respectively.
\begin{equation}
\ket{\psi(t)}=c_1\exp(-iE_1t/\hbar)v_1+c_2\exp(-iE_1t/\hbar)v_2
\label{4.5}
\end{equation}
However, we are not interested in this solution. We are interested in calculating the probability of finding the electron at a particular site. These probabilities are the square of the absolute value of the probability amplitudes $ \phi_{1}(t)$ and $ \phi_{2}(t)$. Most of the time we face an ensemble of such systems. The density matrix fully describes the system and all the observables can be calculated from it. 
\begin{align}
\rho(t)&=\rho_{11}(t)\ket{1}\bra{1}+\rho_{22}(t)\ket{2}\bra{2} \nonumber   \\
       &+\rho_{12}(t)\ket{1}\bra{2}+\rho_{21}(t)\ket{2}\bra{1}
\end{align}
where $\rho_{ij}$ for $i=1,2$ and $j=1,2$ are the matrix elements. Its equation of motion is 
\begin{equation}
\frac{d\rho(t)}{dt}=\frac{-i}{\hbar}(\hat{H}\hat{\rho}-\hat{\rho}\hat{H})
\label{4.6}
\end{equation}
It follows that 
\begin{align}
\rho_{11}(t)&=\frac{1}{2}[\rho_{11}(0)+\rho_{22}(0)        \nonumber  \\
            &+(\rho_{11}(0)-\rho_{22}(0))\cos\left(\omega_{12}t/\hbar\right) \\
            &-i(\rho_{12}(0)-\rho_{21}(0))\sin\left(\omega_{12}t/\hbar\right)]   \nonumber
\end{align}
\begin{align}
\rho_{12}(t)&=\frac{1}{2}[\rho_{12}(0)+\rho_{21}(0) \nonumber \\
            &+(\rho_{12}(0)-\rho_{21}(0))\cos\left(\omega_{12}t/\hbar\right)]   \\
            &i(\rho_{11}(0)-\rho_{22}(0))\sin\left(\omega_{12}t/\hbar\right)] \nonumber
\end{align}
\begin{equation}
\rho_{21}(t)=\rho^{\dagger}_{12}(t)
\end{equation}
\begin{align}   
\rho_{22}(t)&=\frac{1}{2}[\rho_{11}(0)+\rho_{22}(0)   \nonumber   \\
            &+(\rho_{22}(0)-\rho_{11}(0))\cos\left(\omega_{12}t/\hbar\right)\\
            &+i(\rho_{12}(0)-\rho_{21}(0))\sin\left(\omega_{12}t/\hbar\right)]  \nonumber          
\end{align}
where $\omega_{12}=(E_2-E_1)/\hbar=(2T)/\hbar$. The visibility and coherence can be calculated by substitution of the off-diagonal elements of the previous set of equations in Eq. (\ref{3.17}) and Eq. (\ref{3.4}), respectively. We may define the site predictability, a measure of the particle nature, analogously to the path predictability as $P=|\rho_{11}-\rho_{22}|$ which gives for a single electron on two sites
\begin{eqnarray}
P&=&|(\rho_{11}(0)-\rho_{22}(0))\cos\left(\omega_{12}t/\hbar\right) \\
  &+&i (\rho_{21}(0)-\rho_{12}(0))\sin\left(\omega_{12}t\right)|
\end{eqnarray}

\begin{figure}[h]
\includegraphics[scale=0.75]{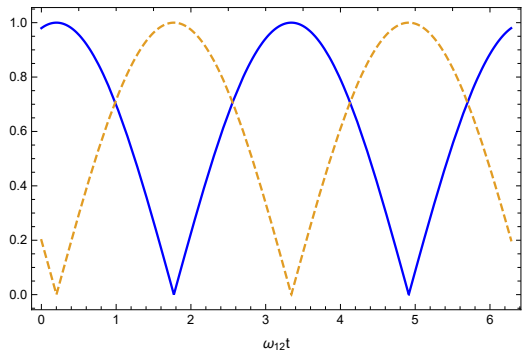}
\caption{Visibility $V_C$ (solid) and site predictability (dashed) for a single electron on two sites using $\phi_1(0)=\sqrt{\frac{60}{100}}$i and $\phi_2(0)=-\sqrt{\frac{40}{100}}$.}
\label{Fig1}
\end{figure}

 If the  system is  in a maximally coherent state, the predictabily, $V_C$ and $V$ are independent of time. They take the values $0$, $1$  and $\frac{1}{3}$, respectively. Figure (\ref{Fig1}) illustrates the variation of the site predictability and  $V_C$  as a function of $\omega_{12}t$. $V_C$ and $P$ are complementary in nature. If one peaks, the other is minimum. The figure was generated for a pure state where the coefficients of the basis states are simple number. That does not have to be the case in realistic situations.

\subsubsection{Two electron on two sites}
The general quantum state for two electrons on two sites can be written in the following form
\begin{align}
\ket{\psi(t)}&=\phi_{1}(t)\ket{\uparrow\downarrow;0}+\phi_{2}(t)\ket{\uparrow;\downarrow} \nonumber \\
             &+\phi_{3}(t)\ket{\downarrow;\uparrow}+\phi_{4}(t)\ket{0;\uparrow\downarrow}
\end{align}
where $\ket{\uparrow\downarrow;0}$, $\ket{\uparrow;\downarrow}$, $\ket{\downarrow;\uparrow}$ and $\ket{0;\uparrow\downarrow}$ are the basis states and $\phi_{1}(t)$ to $\phi_{4}(t)$ are time dependent coefficients in the Schroedinger's picture. 
The density matrix for this system is $4\times4$. Now we solve the equation of motion under the following approximations \\
case I: T=0, the off-diagonal elements evolves as 
\begin{eqnarray}
\rho_{12}(t)&=&\rho_{12}(0)\exp\left(-iUt/\hbar\right) \\
\rho_{13}(t)&=&\rho_{13}(0)\exp\left(-iUt/\hbar\right) \\
\rho_{14}(t)&=&\rho_{14}(0)\exp\left(iUt/\hbar\right) \\
\rho_{23}(t)&=&\rho_{23}(0) \\
\rho_{24}(t)&=&\rho_{24}(0)\exp\left(iUt/\hbar\right) \\
\rho_{34}(t)&=&\rho_{12}(0)\exp\left(iUt/\hbar\right)
\end{eqnarray}
In this approximation, the visibility is time independent since the absolute values of the density matrix elements do not evolve with time. In other words, electron-electron interaction has no effect on the wave properties.\\
Case II: U=0, the off-diagonal elements evolves as 

\begin{align}
\rho_{12}(t)&=\frac{1}{16} [2A_{12}+2B_{12} e^{-i\omega_{12}t}+2C_{12} e^{i\omega_{12}t} \nonumber \\
            &+D_{12} e^{-2i\omega_{12}t}+2E_{12} e^{2i\omega_{12}t}]
\label{Eq4.1}
\end{align}
\begin{align}
\rho_{13}(t)&=\frac{1}{16} [2A_{13}+2B_{12} e^{i\omega_{12}t}+C_{13} e^{-2i\omega_{12}t} \nonumber \\
            &+2D_{13} e^{i\omega_{12}t}+E_{13} e^{2i\omega_{12}t}]
\end{align}
\begin{align}
\rho_{14}(t)&=\frac{1}{16} [2A_{14}+2B_{14} e^{-i\omega_{12}t}+2C_{14} e^{i\omega_{12}t} \nonumber \\
            &+D_{14} e^{-2i\omega_{12}t}+E_{14} e^{2i\omega_{12}t}]
\end{align}
\begin{align}
\rho_{23}(t)&=\frac{1}{8} [A_{23}+B_{23} e^{-i\omega_{12}t}+C_{23} e^{i\omega_{12}t} \nonumber \\
            &+D_{23} e^{2i\omega_{12}t}]
\end{align}
\begin{align}
\rho_{24}(t)&=\frac{1}{16} [A_{24}+B_{24} e^{-i\omega_{12}t}+C_{24} e^{i\omega_{12}t} \nonumber \\
            &+D_{24} e^{2i\omega_{12}t}+E_{24} e^{i\omega_{12}t}+F_{24} e^{2i\omega_{12}t}]
\end{align}
\begin{align}
\rho_{34}(t)&=\frac{1}{16} [2A_{34}+2B_{34} e^{-i\omega_{12}t}+2C_{34} e^{i\omega_{12}t} \nonumber \\
            &+D_{34} e^{-2i\omega_{12}t}+E_{34} e^{2i\omega_{12}t}]
\label{Eq4.6}
\end{align}

\begin{figure}[h]
\includegraphics[scale=0.75]{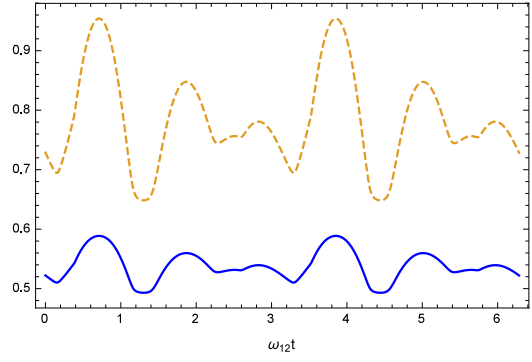}
\caption{ The vertical axis represents $V $ (solid) and $V_C$ (dashed) and the horizontal axis is $\omega_{12}t$ for an ensemble of two electrons on two sites using $\phi_1(0)=\frac{1}{4}+\frac{1}{4}i$, $\phi_2(0)=\frac{1}{4}+\frac{\sqrt{3}}{4}i$, $\phi_3(0)=-\frac{1}{4}$ and $\phi_4(0)=\frac{1}{2}-\frac{\sqrt{5}}{4}i$}
\label{Fig2}
\end{figure}
where the time independent constants $A_{kl}, B_{kl}, C_{kl}, D_{kl}, E_{kl}$ and $F_{kl}$ for $k\neq l$ can be determined from the the initial conditions. Figure (\ref{Fig2}) shows that $V_C$ and $V$ follow the same pattern, but they are shifted with respect to one another. Both of them are independent of the single particle energy $\epsilon_{i}$. Coherence consists of addition of absolute-valued sums of harmonic functions. The frequency of these functions are integer multiples of a fundamental frequency $\omega_{12}=2T/\hbar$. These integers run from $-2$ to $2$. Therefore,  coherence which is the resultant function of the sum is still periodic with a resultant time period equal to $\frac{\pi}{2}\frac{\hbar}{T}$.

\section{Discussion}
Visibility as a measure of the wave nature was introduced. In doing so, the visibility was defined in terms of the extrema of the density matrix in the general case. The minimum density matrix is for incoherent states and the maximum is for maximally coherent states. This definition is an extension of Eq. (\ref{2.4}). This equation was motivated by setting the minimum probability $\abs{\psi}^2_{min}$ to $\abs{\psi_1}^2+\abs{\psi_2}^2$ which is corresponding to a relative phase difference of $\frac{\pi}{2}$. However, one could argue that a phase difference of $\pi$ would yield a smaller probability $\abs{\psi_1}^2+\abs{\psi_2}^2-2\abs{\psi_1 \psi_2}$. If no interference of quantum particles is observed, the square modulus of the resulting wave amplitude is the sum of the square modulus of the constituting components \cite{Gennaro2001} which sets the point of reference to a point of pure particle behavior and justifies the former choice. The latter choice assign the minimum probability amplitude to a point of total destructive interference. For example, if $\psi_{1}=\psi_{2}$, then the probability of finding the quantum at a certain position is zero i.e, total destructive interference not just descruction of the interference terms.

The relation between visibility and a coherence quantifier was presented. We discussed the difference between $V_C$ which is scaled with $n-1$ and $V$ which is scaled with the sum of $\rho^C$ and $\rho^I$. The latter can be used to compare the degree of coherence for different systems. Coherence depends strongly on the intitial values of the density matrix elements as Eq.(\ref{Eq4.1}) to Eq.(\ref{Eq4.6}) show. Site predictability, for two dimensional space, was defined exactly as the path predictability. For higher dimension, $ n$-path predictability needs to be tested for its applicability to systems other than $n$-path interferometers. 
 
\section{Acknowledgments}
I would like to thank the staff and my colleagues in the online course "Applications of Quantum Mechanics", 8.06X, at \url{ https://www.edx.org/course/applications-of-quantum-mechanics} for reading and commenting on this paper.

\end{document}